\documentstyle[12pt,epsf]{article}

  \def\f{\hfill}
\def\s{\hskip2mm} \def\ss{\hskip5mm}

scaled\magstep4 
 
scaled\magstep2 
 
scaled\magstep1 
 
scaled\magstephalf  
  
   \font\sixi=cmmi6
\skewchar\sixi='177 \newbox\hdbox \newcount\hdrows
\newcount\multispancount \newcount\ncase \newcount\ncols \newcount\nrows
\newcount\nspan \newcount\ntemp \newdimen\hdsize \newdimen\newhdsize
\newdimen\parasize \newdimen\spreadwidth \newdimen\thicksize
\newdimen\thinsize \newdimen\tablewidth \newif\ifcentertables
\newif\ifendsize \newif\iffirstrow \newif\iftableinfo \newtoks\dbt
\newtoks\hdtks \newtoks\savetks \newtoks\tableLETtokens
\newtoks\tabletokens \newtoks\widthspec \tableinfotrue \catcode`\@=11
 \def\tstrut{\vrule height3.1ex
depth1.2ex width0pt} \def\and{\char`\&} \def\tablerule{\noalign{\hrule
height\thinsize depth0pt}} \thicksize=1.5pt \thinsize=0.6pt
\def\thickrule{\noalign{\hrule height\thicksize depth0pt}}
  \def\ctr#1{\hfil\ #1\hfil}
  \tablewidth=-\maxdimen
\spreadwidth=-\maxdimen \def\tabskipglue{0pt plus 1fil minus 1fil}
\centertablestrue 
\gdef\ARGS{########} \gdef\headerARGS{####} \def\@mpersand{&}
{\catcode`\|=13 \gdef\letbarzero{\let|0} \gdef\letbartab{\def|{&&}}
\gdef\letvbbar{\let\vb|} } {\catcode`\&=4 \def\ampskip{&\omit\hfil&}
\catcode`\&=13 \let&0 \xdef\letampskip{\def&{\ampskip}}
\gdef\letnovbamp{\let\novb&\let\tab&} } \def\begintable{ \begingroup
\catcode`\|=13\letbartab\letvbbar \catcode`\&=13\letampskip\letnovbamp
\def\multispan##1{ \omit \mscount##1
\multiply\mscount\tw@\advance\mscount\m@ne \loop\ifnum\mscount>\@ne
\sp@n\repeat } \def\|{ &\omit\widevline& } \ruledtable }
\long\def\ruledtable#1\endtable{ \offinterlineskip \tabskip 0pt
\def\widevline{\vrule width\thicksize}
\def\endrow{\@mpersand\omit\hfil\crnorm\@mpersand}
\def\crthick{\@mpersand\crnorm\thickrule\@mpersand}
\def\crthickneg##1{\@mpersand\crnorm\thickrule
\noalign{\vskip-##1}\@mpersand} \def\crnorule{\@mpersand\crnorm\@mpersand}
\def\crnoruleneg##1{\@mpersand\crnorm\noalign{\vskip-##1}\@mpersand}
\let\nr=\crnorule \def\endtable{\@mpersand\crnorm\thickrule}
\let\crnorm=\cr \edef\cr{\@mpersand\crnorm\tablerule\@mpersand}
\def\crneg##1{\@mpersand\crnorm\tablerule \noalign{\vskip-##1}\@mpersand}
\let\ctneg=\crthickneg \let\nrneg=\crnoruleneg \the\tableLETtokens
\tabletokens={&#1} \countROWS\tabletokens\into\nrows
\countCOLS\tabletokens\into\ncols \advance\ncols by -1 \divide\ncols by 2
\advance\nrows by 1 \iftableinfo \immediate\write16{(Nrows=\the\nrows,
Ncols=\the\ncols)} \fi \ifcentertables \ifhmode \par\fi \@@line{ \hss
\else \hbox{ \fi \vbox{ \makePREAMBLE{\the\ncols} \edef\next{\preamble}
\let\preamble=\next \makeTABLE{\preamble}{\tabletokens} } \ifcentertables
\hss}\else }\fi \endgroup \tablewidth=-\maxdimen \spreadwidth=-\maxdimen }
\def\makeTABLE#1#2{ { \let\ifmath0 \let\header0 \let\multispan0 \ncase=0
\ifdim\tablewidth>-\maxdimen \ncase=1\fi \ifdim\spreadwidth>-\maxdimen
\ncase=2\fi \relax \ifcase\ncase \widthspec={} \or
\widthspec=\expandafter{\expandafter t\expandafter o \the\tablewidth}
\else \widthspec=\expandafter{\expandafter s\expandafter p\expandafter r
\expandafter e\expandafter a\expandafter d \the\spreadwidth} \fi
\xdef\next{ \halign\the\widthspec{ #1 \noalign{\hrule height\thicksize
depth0pt} \the#2\endtable } } } \next } \def\makePREAMBLE#1{ \ncols=#1
\begingroup \let\ARGS=0 \edef\xtp{\widevline\ARGS\tabskip\tabskipglue
&\ctr{\ARGS}\tstrut} \advance\ncols by -1 \loop \ifnum\ncols>0
\advance\ncols by -1 \edef\xtp{\xtp&\vrule
width\thinsize\ARGS&\ctr{\ARGS}} \repeat
\xdef\preamble{\xtp&\widevline\ARGS\tabskip0pt \crnorm} \endgroup }
\def\countROWS#1\into#2{ \let\countREGISTER=#2 \countREGISTER=0
\expandafter\ROWcount\the#1\endcount } \def\ROWcount{
\afterassignment\subROWcount\let\next= } \def\subROWcount{
\ifx\next\endcount \let\next=\relax \else \ncase=0 \ifx\next\cr
\global\advance\countREGISTER by 1 \ncase=0 \fi \ifx\next\endrow
\global\advance\countREGISTER by 1 \ncase=0 \fi \ifx\next\crthick
\global\advance\countREGISTER by 1 \ncase=0 \fi \ifx\next\crnorule
\global\advance\countREGISTER by 1 \ncase=0 \fi \ifx\next\crthickneg
\global\advance\countREGISTER by 1 \ncase=0 \fi \ifx\next\crnoruleneg
\global\advance\countREGISTER by 1 \ncase=0 \fi \ifx\next\crneg
\global\advance\countREGISTER by 1 \ncase=0 \fi \ifx\next\header \ncase=1
\fi \relax \ifcase\ncase \let\next\ROWcount \or \let\next\argROWskip \else
\fi \fi \next } \def\counthdROWS#1\into#2{ \dvr{10} \let\countREGISTER=#2
\countREGISTER=0 \dvr{11} \dvr{13} \expandafter\hdROWcount\the#1\endcount
\dvr{12} } \def\hdROWcount{ \afterassignment\subhdROWcount\let\next= }
\def\subhdROWcount{ \ifx\next\endcount \let\next=\relax \else \ncase=0
\ifx\next\cr \global\advance\countREGISTER by 1 \ncase=0 \fi
\ifx\next\endrow \global\advance\countREGISTER by 1 \ncase=0 \fi
\ifx\next\crthick \global\advance\countREGISTER by 1 \ncase=0 \fi
\ifx\next\crnorule \global\advance\countREGISTER by 1 \ncase=0 \fi
\ifx\next\header \ncase=1 \fi \relax \ifcase\ncase \let\next\hdROWcount
\or \let\next\arghdROWskip \else \fi \fi \next } {\catcode`\|=13\letbartab
\gdef\countCOLS#1\into#2{ \let\countREGISTER=#2 \global\countREGISTER=0
\global\multispancount=0 \global\firstrowtrue
\expandafter\COLcount\the#1\endcount \global\advance\countREGISTER by 3
\global\advance\countREGISTER by -\multispancount } \gdef\COLcount{
\afterassignment\subCOLcount\let\next= } {\catcode`\&=13
\gdef\subCOLcount{ \ifx\next\endcount \let\next=\relax \else \ncase=0
\iffirstrow \ifx\next& \global\advance\countREGISTER by 2 \ncase=0 \fi
\ifx\next\span \global\advance\countREGISTER by 1 \ncase=0 \fi \ifx\next|
\global\advance\countREGISTER by 2 \ncase=0 \fi \ifx\next\|
\global\advance\countREGISTER by 2 \ncase=0 \fi \ifx\next\multispan
\ncase=1 \global\advance\multispancount by 1 \fi \ifx\next\header \ncase=2
\fi \ifx\next\cr \global\firstrowfalse \fi \ifx\next\endrow
\global\firstrowfalse \fi \ifx\next\crthick \global\firstrowfalse \fi
\ifx\next\crnorule \global\firstrowfalse \fi \ifx\next\crnoruleneg
\global\firstrowfalse \fi \ifx\next\crthickneg \global\firstrowfalse \fi
\ifx\next\crneg \global\firstrowfalse \fi \fi \relax \ifcase\ncase
\let\next\COLcount \or \let\next\spancount \or \let\next\argCOLskip \else
\fi \fi \next } \gdef\argROWskip#1{ \let\next\ROWcount \next }
\gdef\arghdROWskip#1{ \let\next\ROWcount \next } \gdef\argCOLskip#1{
\let\next\COLcount \next } } } \def\spancount#1{ \nspan=#1\multiply\nspan
by 2\advance\nspan by -1 \global\advance \countREGISTER by \nspan
\let\next\COLcount \next} \def\dvr#1{\relax} \def\header#1{
\dvr{1}{\let\cr=\@mpersand \hdtks={#1} \counthdROWS\hdtks\into\hdrows
\advance\hdrows by 1 \ifnum\hdrows=0 \hdrows=1 \fi
\dvr{5}\makehdPREAMBLE{\the\hdrows} \dvr{6}\getHDdimen{#1}
{\parindent=0pt\hsize=\hdsize{\let\ifmath0
\xdef\next{\valign{\headerpreamble #1\crnorm}}}\dvr{7}\next\dvr{8} }
}\dvr{2}} \def\makehdPREAMBLE#1{ \dvr{3} \hdrows=#1 { \let\headerARGS=0
\let\cr=\crnorm \edef\xtp{\vfil\hfil\hbox{\headerARGS}\hfil\vfil}
\advance\hdrows by -1 \loop \ifnum\hdrows>0 \advance\hdrows by -1
\edef\xtp{\xtp&\vfil\hfil\hbox{\headerARGS}\hfil\vfil} \repeat
\xdef\headerpreamble{\xtp\crcr} } \dvr{4}} \def\getHDdimen#1{ \hdsize=0pt
\getsize#1\cr\end\cr } \def\getsize#1\cr{ \endsizefalse\savetks={#1}
\expandafter\lookend\the\savetks\cr \relax \ifendsize \let\next\relax
\else \setbox\hdbox=\hbox{#1}\newhdsize=1.0\wd\hdbox
\ifdim\newhdsize>\hdsize \hdsize=\newhdsize \fi \let\next\getsize \fi
\next } \def\lookend{\afterassignment\sublookend\let\looknext= }
\def\sublookend{\relax \ifx\looknext\cr \let\looknext\relax \else \relax
\ifx\looknext\end \global\endsizetrue \fi \let\looknext=\lookend \fi
\looknext } \def\tablelet#1{
\tableLETtokens=\expandafter{\the\tableLETtokens #1} } \catcode`\@=12
\newcommand{\dmsq}{$\Delta m^2$} \newcommand{\etal}{{\em et al.}}
\newcommand{\nuebar} {$\overline{\nu}_{e}$} \newcommand{\C}{$\,^{\circ}$C}
\def\SN#1E#2 {\mbox{$#1\times10^{#2}$}} \def\isotope#1{\mbox{${}^{#1}$}}
\def\Reines{$\overline{\nu}_e\,+\,p\,\rightarrow\,e^+\,+\,n$}
 \def\r{\hskip.5mm} \def\units#1{\hbox{$\,{\rm #1}$}}

\begin{document}
\begin{flushright}
DUHEP-9306001\\
June 1993\\
\end{flushright}
\vspace*{5mm}
\begin{center}
 {\bf
        CHOOZ AND PERRY: NEW EXPERIMENTS FOR  \\
        LONG BASELINE REACTOR NEUTRINO OSCILLATIONS}
\footnote
{Presented at the 5th International Workshop on Neutrino Telescopes,
Venice, March 1993, on behalf of the Chooz Collaboration:
{\em Coll\`ege de France:} H.~deKerret, D.~Kryn, B.~Lefi\`evre,
     M.~Obolensky, S.~Sukhotin, P.~Courty and D.~Marchand;
{\em Drexel University:} C.E.~Lane, R.~Steinberg, and F.C.~Wang;
{\em Kurchatov Institute:} V.P.~Martemyanov, L.A.~Mikaelyan,
     M.D.~Skorokhvatov, and V.N.~Vyrodov;
{\em LAPP (Annecy):} Y.~Declais, J.~Favier and A.~Oriboni;
{\em Pisa University:} A.~Baldini, C.~Bemporad, F.~Cei, M.~Grassi,
                                        D.~Nicol\'o and R.~Pazzi;
{\em Trieste University:}  G.~Giannini;
{\em Univ.~of California (Irvine):} Z.D.~Greenwood, W.R.~Kropp, L.~Price,
     S.~Riley and H.~Sobel;
{\em Univ.~of New Mexico:} B.B.~Dieterle and R.~Reeder.}

\vspace{4mm}
R.I. Steinberg

\vspace{2mm}
{\em Department of Physics, Drexel University, Philadelphia, PA  19104 \\
(steinberg@duphy4.drexel.edu)}

\vspace{4mm}
\end{center}

\begin {abstract}

We discuss the Chooz experiment, a long baseline search for neutrino
vacuum oscillations, which will utilize a gadolinium-loaded liquid
scintillation detector one km from a large nuclear power station. The
300-mwe underground site of the detector reduces cosmic ray muons, the
main source of background in this type of experiment, by a factor of 300,
thereby allowing clean detection of antineutrinos from the reactor. The
experimental goal is to probe \dmsq\ values down to \SN1E-3 \units{eV^2}
for large values of ${\rm sin}^2\,2\theta$ and mixing angles to 0.08 for
favorable regions of \dmsq. A subsequent experiment which will have a 13
km baseline at the former IMB site in Ohio and which can reach $\Delta m^2
\ge \SN8E-5 \units{eV^2}$ is also briefly described.

\end{abstract}
\newpage

\section  {Introduction}

A central issue in particle physics, astrophysics and cosmology is the
question of whether or not the rest masses of the neutrinos are exactly
zero. In the minimal $SU(2)_L\otimes U(1)$ standard electroweak model, all
neutrinos are massless and lepton number is exactly conserved. Despite
longstanding success, however, the standard model is incomplete and
inadequate, with many parameters left unspecified and a physically
unreasonable global symmetry needed to enforce exact lepton number
conservation.

The standard model, therefore, generally is agreed to need significant
extensions. Although the nature of the necessary modifications is unclear,
most proposed extensions of the standard model allow finite neutrino mass
and many others require it. Experimental searches for neutrino mass are
therefore important both to test the standard model and to guide theorists
seeking a better model. Clear experimental evidence for finite neutrino
mass would herald a new era of physics ``beyond the standard model''.

Furthermore, if the neutrinos are experimentally proven to have mass, not
only would there be deep implications for our theoretical ideas on
particle physics and on the unification of forces, but also for our
understanding of essential astrophysical phenomena such as the energy
generating mechanism of the Sun, the final stages of stellar evolution,
and the history of the Universe itself. Thus, the experimental search for
finite neutrino mass could well provide the key to a rich domain of new
phenomena.

We discuss here an incisive new experiment combining and extending
proven methods of particle detection and background rejection to search
with order-of-magnitude improved sensitivity for the phenomenon of
neutrino oscillations. If discovered, such oscillations would provide
clear proof of the existence of finite neutrino mass.

This area is ripe for a new and potentially definitive experiment, in that
experimental hints for the existence of neutrino oscillations come from
two distinct sets of observations: a long series of unexpectedly low
results in measurements of the terrestrial flux of solar neutrinos; and
the observation of an anomaly in the flux of cosmic-ray produced muon
neutrinos observed by deep underground detectors. Achieving sensitivity to
the neutrino mass scales indicated by these hints requires a new
generation of long baseline neutrino oscillation experiments such as the
Chooz experiment.

\section  {Reactor Neutrino Oscillation Experiments}

Nuclear power reactors provide intense, well-understood sources of
low energy \nuebar 's. Not only is the neutrino energy spectrum known to
an accuracy of a few percent, but the neutrino flavor composition (pure
\nuebar ) and angular distribution (isotropic) of the reactor neutrino
beam are known essentially perfectly, unlike accelerator neutrino beams.
Thus, reactors provide nearly ideal sources for neutrino oscillation
experiments.

Over the years, therefore, a number of neutrino oscillation searches have
been performed. Table~\ref{t:old_experiments} compares some of these
experiments, as well as three future experiments. The best existing limits
from a reactor neutrino oscillation measurement come from an experiment at
G\"osgen in Switzerland which was 65 meters from the reactor and achieved
a \dmsq\ sensitivity approaching $\sim10^{-2}$\units{eV^2} \cite{Zacek}. A
more stringent limit in \dmsq\ has been obtained recently by a Kurchatov
Institute group \cite{Vidyakin}, but the sensitivity in mixing was not as
good (${\rm sin}^2\,2\theta\,\sim.50$).

The G\"osgen experiment measured the \nuebar\ spectrum at three distances
up to 64.7\units{m} from the reactor by using the inverse beta decay
reaction \Reines. The reactor had a thermal power of 2800\units{MW}, while
the detector provided a target mass of 320\units{kg} and yielded a neutron
detection efficiency of 21.7\%. The results of the G\"osgen experiment as
well as several other reactor neutrino oscillation experiments are
summarized on the exclusion plot shown in Fig.~\ref{f:reactor}, where, for
example, the area above and to the right of the curve labelled
``G\"osgen'' is the region of the \dmsq\ vs.~${\rm sin}^2\,2\theta$ phase
space where neutrino oscillations would have been detected had they been
present. The minimum detectable \dmsq\ for maximal mixing was \SN1.9E-2
\units{eV^2}.

\section  {Description of the Chooz Experiment}

The Chooz experiment is a highly sensitive neutrino oscillation search
using an underground facility near a nuclear power station in France. The
experiment will look for the flux reduction and spectral distortion which
would signal the existence of vacuum oscillations of the electron
antineutrino beam emitted by the reactors. The limits on neutrino
oscillations expected are shown in Fig.~\ref{f:reactor} by the curves
labelled ``Chooz''. For maximal mixing, the experiment will detect
oscillations for values of \dmsq $>$ \SN1E-3 \units{eV^2}, an
order-of-magnitude improvement over currently available limits. The
experiment will also provide an excellent testing ground for methodology
for the future Perry experiment (see Table~\ref{t:old_experiments} and
ref.~\cite{Perry}). An overview of the Chooz experiment is shown in
Fig.~\ref{f:chovervw}.

\section  {Neutrino Source and Site}

The neutrino source is a pair of reactors at the Chooz B nuclear power
station in the Ardennes region of northeastern France. Each of the PWR
reactors will have a thermal power of 4.2\units{GW} and is scheduled for
startup by the end of 1995. An essential feature of the experimental site
is the availability of a tunnel with an overburden of 115\units{m} of
rock, equivalent to 300\units{m} of water. Building the neutrino detector
in this tunnel will provide the cosmic-ray shielding needed to preserve
the signal/noise ratio against a one hundred-fold neutrino flux reduction
with respect to previous experiments. The advantages of such a site are
made clear in Fig.~\ref{f:mu_nu_di}.

\section  {The Detector}

The neutrino detector is shown in Fig.~\ref{f:choozdet}, while its
principal design parameters are exhibited in Table~\ref{t:chdespar}. The
neutrino target will be contained in a 5.5-m-diameter cylindrical steel
tank shielded locally by about 75\units{cm} of low radioactivity material.
The tank will contain three concentric liquid scintillation detectors: an
outer 90-ton veto counter; an intermediate 17-ton optically separated
event containment detector; and a central acrylic vessel containing five
tons of a specially developed gadolinium-loaded liquid scintillator. The
outer two vessels will contain a high flash point pure hydrocarbon
scintillator also specially developed at Drexel for this experiment.

Scintillation photons from particle interactions in the two inner
detectors will be collected by 160 eight-inch photomultiplier tubes and
processed by fast multi-hit TDC's and fast waveform digitizers. The
detector will have good energy resolution, with about 76 photoelectrons
detected per MeV of ionization energy deposited. Scintillation light from
the veto counter will be collected by an additional 40 PMT's.

Primary shielding against background will be provided by the
300\units{mwe}-thick rock overburden, which will reduce the surface cosmic
ray muon flux by a factor of 300, as shown in Fig.~\ref{f:mu_nu_di}. The
residual cosmic ray background is further suppressed by the 90-ton outer
veto counter, which also provides additional shielding against ambient
radioactivity.

Neutrinos above the threshold energy of 1.8\units{MeV} will be detected by
the reaction \Reines. The observable energy from this
reaction will equal the positron kinetic energy augmented by
1.022\units{MeV} resulting from detection of the positron annihilation
gamma rays. Following thermalization of the recoil neutron, an additional
8\units{MeV} will be detected as a result of capture of the neutron by a
gadolinium nucleus (time constant 28\units{\mu s}). Thus, a readily
recognizable delayed coincidence pulse pair will signal the neutrino
interaction. The 8-MeV neutron capture event will be well separated
from the beta and gamma radiation accompanying decay of members of the
ubiquitous uranium and thorium decay chains and from the decay of
\isotope{40}K. Further significant rejection of the accidental coincidence
background will be possible by software reconstruction of the positron and
neutron capture vertices.

With an expected event rate of about 31\units{d^{-1}} (see
Table~\ref{t:choozrat}), we anticipate a total running time of two years
will be adequate to achieve our goal of a statistical error better than
4\%. Neutrino oscillations would be uncovered by comparing the observed
integral count rate with an accurately calibrated detector-efficiency
Monte Carlo calculation. Information on reactor power and fuel burn-up
will be available from the power station. The neutrino flux per unit power
will be determined from calculations which currently are reliable to about
2\%, especially at low energy, where a neutrino oscillation signal would
probably be strongest. In the event of a positive signal, it would be
useful to construct a close-in detector to provide a flux independent
confirmation. A detector site at less than 100\units{m} from the reactors
is available if needed.

For the experiment, about five tons of Gd-loaded scintillator and 107~tons
of high flash point hydrocarbon scintillator will be needed. Optimized
formulations for each of these materials were developed after extensive
testing at Drexel University. We thus have high performance detection
media for the experiment at minimal cost. The properties of the two
scintillators are listed in Table~\ref{t:gadcomp}. Fig.~\ref{f:ncap9305}
shows a measurement of the neutron capture time spectrum for the Gd-loaded
scintillator. Similar measurements over more than a one year period show
no changes in the mean capture lifetime, indicating good long-term
stability of the material.

\section  {Conclusions and Outlook}

The availability of a site near a reactor yet well-shielded from cosmic
rays provides a new opportunity to extend the search for neutrino
oscillations by an order of magnitude. The Chooz experiment described here
will cover the region of parameter space hinted at by the anomalous
atmospheric neutrino results. The experiment can be running by mid-1995,
with preliminary results in hand before the end of 1996.

Extension of sensitivity by an additional order of magnitude in \dmsq\
will be possible in a future experiment at the Perry reactor using the
deep underground site of the former IMB experiment \cite{Perry}. Many of
the methods and measurements of the Chooz experiment will be directly
applicable to Perry. The Perry experiment would have a fiducial volume of
one kiloton and a count rate of 12\units{d^{-1}}.


\begin{table}[htbp]
\def\tstrut{\vrule height 2.00ex depth .80ex width 0pt}
\thicksize=1.pt\thinsize=.4pt\begintable\tstrut
\f EXPERIMENT    \f|\r NEUTRINO\r|\f REACTOR  \f|\f\dmsq       \f\nr
\r               \f|\f TARGET  \f|\r DISTANCE \r|\f(min.)      \f\nr
\r               \f|\f MASS    \f|\f (max.)   \f|\r(90\% c.l.) \f\crthick
\r Grenoble      \f|\f  320 kg\f|\f  8.75 m  \f|\f .15          \f\nr
\r Savannah River\r|\f  260 kg\f|\f   24 m   \f|\f .05          \f\nr
\r G\"osgen      \f|\f  320 kg\f|\f   65 m   \f|\f .019         \f\nr
\r Krasnoyarsk   \f|\f  600 kg\f|\f   92 m   \f|\f .014$^*$     \f\nr
\r Rovno         \f|\f  200 kg\f|\f   25 m   \f|\f .06          \f\nr
\r Krasnoyarsk   \f|\f  600 kg\f|\f  230 m   \f|\f .01          \f\nr
\r Bugey III     \f|\f 1200 kg\f|\f   40 m   \f|\f in progress  \f\nr
\r San Onofre    \f|\f12000 kg\f|\f  680 m   \f|\f \SN2E-3      \f\nr
\r Chooz         \f|\f 4800 kg\f|\f 1025 m   \f|\f \SN1E-3      \f\nr
\r Perry         \f|\f1000 ton\f|\f 12.9 km  \f|\f \SN1E-4      \f\nr
\r               \f|\r        \f|\f          \f|\f$^*$68\% c.l.  \f\endtable
\bigskip
\caption[Summary of Reactor Neutrino Oscillation Experiments]
{Summary of reactor neutrino oscillation experiments.}
\label{t:old_experiments}
\end{table}

\begin{table}[t]
\epsfysize=2.75in
\centerline{\epsffile[ 81 297 522 544]{c:/papers/ps/chdespar.eps}} 
\caption[Design Parameters of the Detector]
{Design parameters of the detector.}
\label{t:chdespar}
\end{table}

\begin{table}[t]
\epsfysize=3.25in
\centerline{\epsffile[ 81 207 540 585]{c:/papers/ps/choozrat.eps}} 
\caption[Neutrino Event Rate Calculation]
{Calculation of the neutrino event rate.}
\label{t:choozrat}
\end{table}

\begin{table}[b]
\def\tstrut{\vrule height 2.00ex depth .80ex width 0pt}
\thinsize=.4pt \thicksize=1.pt \begintable
\s                               \f|\f {\bf Gadolinium}           \f|\f   {\bf
High Flash}         \f\crthick
\s  Chemical content:            \f|\f                            \f|\f
                   \f\nr
\ss mineral oil                  \f|\f59.5\%                      \f|\f96.8\%
                   \f\nr
\ss aromatics and alcohols       \f|\f40\%                        \f|\f 3\%
                   \f\nr
\ss PPO, bis-MSB, etc.           \f|\f0.4\%                       \f|\f0.2\%
                   \f\nr
\ss Gd                           \f|\f0.1\%                       \f|\f  -
                   \f\cr
\s  Compatibility                \f|\f\s acrylic, ABS, PVC, CF4\s \f|\f\s
acrylic, ABS, PVC, CF4\s \f\cr
\s  Density (25\C)               \f|\f0.869 gm cm$^{-3}$          \f|\f0.854 gm
cm$^{-3}$          \f\cr
\s  Volume expansion coeff.      \f|\f\SN7.9E-4  K$^{-1}$
\f|\f\SN7.4E-4  K$^{-1}$         \f\cr
\s  H/C ratio                    \f|\f1.93                        \f|\f2.07
                   \f\cr
\s  Scintillation yield          \f|\f42\% of anthracene          \f|\f42\% of
anthracene          \f\nr
\s                               \f|\f{\em -or-} 162 eV/photon    \f|\f{\em
-or-} 162 eV/photon    \f\cr
\s  Optical attenuation length   \f|\f8 m                         \f|\f20 m
                   \f\cr
\s  Refractive index (20\C)      \f|\f  1.480$\pm$.002            \f|\f
1.473$\pm$.002             \f\cr
\s  Flash point                  \f|\f 69\C                       \f|\f 110\C
                   \f\cr
\s  Atomic composition:          \f|\f                            \f|\f
                   \f\nr
\f      H                        \f|\f\SN7.00E22 \,atoms
cm$^{-3}$\f|\f\SN7.57E22 \,atoms cm$^{-3}$\f\nr
\f      C                        \f|\f\SN3.62E22 \,atoms
cm$^{-3}$\f|\f\SN3.65E22 \,atoms cm$^{-3}$\f\nr
\f      N                        \f|\f\SN1.41E19 \,atoms
cm$^{-3}$\f|\f\SN4.01E18 \,atoms cm$^{-3}$\f\nr
\f      O                        \f|\f\SN1.08E21 \,atoms
cm$^{-3}$\f|\f\SN4.01E18 \,atoms cm$^{-3}$\f\nr
\f      Gd                       \f|\f\SN3.33E18 \,atoms cm$^{-3}$\f|\f   -
                   \f\cr
\s Neutron capture time          \f|\f  28\,$\mu$s                \f|\f
180\,$\mu$s                \f\cr
\s Gd capture fraction           \f|\f  87\%                      \f|\f  -
                   \f\cr
\s Thermal $n$ mean capture path \f|\f  6\,cm                     \f|\f  40\,cm
                   \f\endtable
\caption[Liquid Scintillator Properties]
{Properties of the liquid scintillators developed at Drexel for
use in the experiment.}
\label{t:gadcomp}
\end{table}
\clearpage

\begin{figure}[htbp]
\epsfysize=4in
\caption[Reactor Neutrino Exclusion Plot]
{Reactor neutrino oscillation limits showing 90\% c.l.~exclusion contours.
The Chooz experiment will extend existing results by more than one order
of magnitude in \dmsq. The future Perry experiment would allow an
additional ten-fold sensitivity improvement.}
\label{f:reactor}
\end{figure}

\begin{figure}[htbp]
\epsfysize=5in
\centerline{\epsffile[136 120 500 700]{c:/papers/ps/chovervw.eps}} 
\caption[Overview of the Experiment]
{Overview of the experiment. Detection of the weak neutrino signal at
1\units{km} is made possible by the 300 meter of water equivalent
shielding above the detector, which attenuates the otherwise overwhelming
cosmic ray muon-induced background by a factor of 300 and by the layered
design of the detector and its local shielding. The high detection
efficiency for antineutrino events (more than 80\%) further enhances the
expected signal-to-background ratio and facilitates accurate determination
of the detector efficiency, needed for accurate comparison of the measured
and expected neutrino fluxes.}
\label{f:chovervw}
\end{figure}

\begin{figure}[htbp]
\epsfysize=4in
\caption[Muon Depth Intensity and Neutrino Flux]
{Muon depth vs.~intensity and neutrino flux at various sites. At Chooz
(and to a somewhat lesser extent at Perry/IMB) the low neutrino fluxes
available for long baseline oscillation experiments are compensated by
comparably reduced muon fluxes. Since cosmic ray muons produce the most
serious backgrounds in these experiments, shallow sites such as that
at San Onofre are at a serious disadvantage.}
\label{f:mu_nu_di}
\end{figure}

\begin{figure}[htbp]
\epsfysize=5in
\centerline{\epsffile[ 72 144 576 720]{c:/papers/ps/choozdet.eps}} 
\caption[The Detector]
{The Chooz detector. The neutrino target contains 4.9 tons of
gadolinium-loaded liquid scintillator in which the reaction \Reines\ takes
place. The positrons and their annihilation photons are detected at zero
time delay, while the neutron is detected typically between 10 and
100\units{\mu s} later following thermalization and capture by a
gadolinium nucleus, leading to the release of a total of about
8\units{MeV} of $\gamma $ rays. A 300 meter of water equivalent
underground location and massive gravel, steel and liquid scintillator
shielding suppress the background to a few counts per day, about 10 times
lower than the anticipated neutrino event rate. }
\label{f:choozdet}
\end{figure}

\begin{figure}[htbp]
\epsfysize=4in
\caption[Gd Neutron Capture Time Spectrum]
{Neutron capture time delay spectrum with a 0.12\% Gd-loaded scintillator.
Thermal neutron capture events produce the exponentially decaying part of
the curve. A mean capture lifetime of
22.5\units{\mu s} is obtained by fitting an exponential plus constant term
to this region. For the Chooz scintillator, a Gd concentration of 0.1\%
thus will yield a capture lifetime of about 27\units{\mu s}. The prompt
peak, indicated by the arrow, has been shifted by delaying the signal from
the neutron channel. Prompt coincidences are caused both by cosmic ray
showers and by scattered gamma rays from the Cm/Be neutron source.}
\label{f:ncap9305}
\end{figure}

\end{document}